\documentclass[aps,preprint]{revtex4}%
\usepackage{amsfonts}
\usepackage{amsmath}
\usepackage{amssymb}
\usepackage{graphicx}%
\begin{document}
\title{Photon-number-resolving decoy state quantum key distribution}
\author{Qing-yu Cai$^{1\ast}$ and Yong-gang Tan$^{1,2}$}
\affiliation{$^{1}$State Key Laboratory of Magnetic Resonances and Atomic and molecular
physics, Wuhan Institute of Physics and Mathematics, The Chinese Academy of
Sciences, Wuhan 430071, P. R. China}
\affiliation{$^{2}$Graduation University of Chinese Academy of Sciences}

\begin{abstract}
In this paper, a photon-number-resolving decoy state quantum key distribution
scheme is presented based on recent experimental advancements. A new upper
bound on the fraction of counts caused by multiphoton pulses is given. This
upper bound is independent of intensity of the decoy source, so that both the
signal pulses and the decoy pulses can be used to generate the raw key after
verified the security of the communication. This upper bound is also the lower
bound on the fraction of counts caused by multiphoton pulses as long as faint
coherent sources and high lossy channels are used. We show that Eve's coherent
multiphoton pulse (CMP) attack is more efficient than symmetric individual
(SI) attack when quantum bit error rate is small, so that CMP attack should be
considered to ensure the security of the final key. finally, optimal intensity
of laser source is presented which provides 23.9 km increase in the
transmission distance.

\end{abstract}

\maketitle

\section{\bigskip introduction}

Quantum key distribution (QKD) is a physically secure method, by which private
key can be created between two partners, Alice and Bob, who share a quantum
channel and a public authenticated channel [1]. The key bits then be used to
implement a classical private key cryptosystem, or more precisely called
$one-time$ $pad$ algorithm, to enable the partners to communicate securely.
The best known QKD is the BB84 protocol published by Bennett and Brassward in
1984 [2], security of which has been studied deeply [3-7].

Experimental BB84 QKD was demonstrated by many groups [8]. An optical BB84 QKD
system includes the photon sources, quantum channels, single-photon detectors,
and quantum random-number generators. In principle, optical quantum
cryptography is based on the use of single-photon Fock states. However,
perfect single-photon sources are difficult to realize experimentally.
Practical implementations rely on weak laser pulses in which photon number
distribution obeys Possionian statistics. Thus, no-cloning principle is
ineffective in the case of multiphoton pulses. If the quantum channel is high
lossy, Eve can obtain full information of the final key by using photon number
splitting (PNS) attack without being detected [9-13]. In GLLP [7], it has been
shown that the secure final key of BB84 protocol can be extracted from sifted
key at the asymptotic rate
\begin{equation}
R=(1-\Delta)-H_{2}(e)-H_{2}(e+\Delta),
\end{equation}
where $e$ is the quantum bit error rate (QBER) found in the verification test
and $\Delta$ is the fraction of counts caused by multiphoton pulses. This
means that both the QBER $e$ and the fraction of tagged signals $\Delta$ are
important to generate the secure final key. It has been shown that Eve's PNS
attack will be limited when Alice and Bob use the decoy-state protocols
[14-20] or the nonorthogonal states scheme [21]. In the decoy-state protocols
[14-20], an important assumption is that the detection apparatus cannot
resolve the photon number of arriving signals. Recently, some
photon-number-resolving detection apparatus were presented [22-24], especially
the noise-free high-efficiency photon-number-resolving detectors [24]. Thus, a
lower upper bound on the fraction of counts $\Delta$ is desired with the
photon-number-resolving detectors. As a matter of fact, Eve's some other
attacks, such as coherent multiphoton pulse (CMP) attack, should also be
considered or else security of the final key will be unreliable.

In this paper, we present a photon-number-resolving decoy state (PDS) quantum
key distribution scheme based on recent experimental advancements. We show
that the upper bound on fraction of counts caused by multiphoton pulses is
$\mu$, no matter how high the channel loss is. We show that coherent
multiphoton pulse (CMP) attack is more efficient than symmetric individual
(SI) attack. We present the optimal approach to generate the sifted key from
the raw key. Optimal parameter of intensity of laser source is presented to
generate the secure final key. This paper is organized as follow: We first
introduce our PDS QKD scheme. Then we discuss Eve's CMP attack. Next, we
present the optimal approach to generate the sifted key from the raw key. Then
we discuss how to select optimal intensity of laser source to generate the
secure final key. Finally, we discuss and conclude.

\section{photon-number-resolving decoy state quantum key distribution}

At present, practical \textquotedblleft single-photon\textquotedblright%
\ sources rely on weak laser pulses in which photon number distribution obeys
Possionian statistics. Most often, Alice sends to Bob a weak laser pulse in
which she has encoded her bit. Each pulse is a priori in a coherent state
$|\sqrt{\mu}e^{i\theta}\rangle$ of weak intensity. Since Eve and Bob have no
information on $\theta$, the state reduces to a mixed state $\rho=\int
\frac{d\theta}{2\pi}|\sqrt{\mu}e^{i\theta}\rangle\langle\sqrt{\mu}e^{i\theta
}|$ outside Alice's laboratory. This state is equivalent to the mixture of
Fock state $\sum_{n}p_{n}|n\rangle\langle n|$, with the number $n$ of photons
distributed as Possionian statistics $p_{n}=p_{\mu}[n]=\mu^{n}e^{-\mu}/n!$.
The source that emits pulses in coherent states $|\sqrt{\mu}e^{i\theta}%
\rangle$ is equivalent to the representation as below: With probability
$p_{0}$, Alice does nothing; With probability $p_{n}$ $(n>0)$, Alice encodes
her bit in $n$ photons. In order to gain Alice's encoding information, Eve
first performs a nondemolition measurement to gain the photon number of the
laser pulses. When she finds there is only one photon in the pulses, she may
implement symmetric individual (SI) attack on this qubit [12]. Otherwise, if
there are two or more than two photons in the pulses, she may implement PNS
attack on Alice's qubit. In long distance QKD, the channel transmittance
$\eta$ can be rather small. If $\eta<(1-e^{-\mu}-\mu e^{-\mu})/\mu$, Eve can
gain full information of Bob's final key by using the PNS attack [11].

In order to detect Eve's PNS attack, Alice can introduce a decoy source
$\mu^{\prime}$ to ensure the security of their QKD. Since Bob's detection
apparatus is sensitive to the photon number, in the absence of Eve, photon
number distributions in Bob's detectors are also Poissonian (Here, we assume
that the dark counts rate $r_{dark}$ in Bob's detectors is zero. We will
discuss the realistic condition of that $r_{dark}>0$ later.),
\begin{align}
p_{sig}^{loss}[n]  &  =\frac{(\eta\mu)^{n}}{n!}e^{(-\eta\mu)},\\
p_{dec}^{loss}[n]  &  =\frac{(\eta\mu^{\prime})^{n}}{n!}e^{(-\eta\mu^{\prime
})}.
\end{align}
Without the decoy state, the necessary condition of that Eve can implement her
PNS attack without being detected is [11]%
\begin{equation}
p_{sig}[n](1-\sum_{i=0}^{n-1}f(n,i))+\sum_{j=n+1}^{\infty}p_{sig}[j]f(j,n)\geq
p_{sig}^{loss}[n],
\end{equation}
where $f(m,k)$ is the probability of that Eve forwards $k$ photons to Bob and
stores the other $m-k$ photons. In general, let us assume Eve implements PNS
attack $P_{n}$ on Alice's pulses. Consider the case of that decoy states are
used by Alice. Essentially, the idea of decoy-state is that [17]
\begin{align}
P_{n}(signal)  &  =P_{n}(decoy)=P_{n}\\
e_{n}(signal)  &  =e_{n}(decoy)=e_{n}.
\end{align}
In this case, Eve can implement her PNS attack without being detected if and
only if that
\begin{align}
p_{sig}[n](1-\sum_{i=0}^{n-1}f(n,i))+\sum_{j=n+1}^{\infty}p_{sig}[j]f(j,n)  &
=p_{sig}^{loss}[n],\\
p_{dec}[n](1-\sum_{i=0}^{n-1}f(n,i))+\sum_{j=n+1}^{\infty}p_{dec}[j]f(j,n)  &
=p_{dec}^{loss}[n].
\end{align}
Using the Taylor series, we can obtain that
\begin{align}
f(n,i)  &  =\binom{n}{i}\eta^{i}(1-\eta)^{n-i},\\
f(j,n)  &  =\binom{j}{n}\eta^{n}(1-\eta)^{j-n}.
\end{align}
Experimentally, these solutions just correspond to the case of that Eve blocks
every photon with the probability $1-\eta$, i.e., Eve forwards every photon
with probability $\eta$ through her lossless channel (This can be realized by
using a beam splitter with the reflection probability $1-\eta$ and the
transmission probability $\eta$.). We will calculate the amount of information
Eve can gain by using her PNS attack described by the equations (7) and (8)
later. \ 

\section{\bigskip coherent multiphoton pulse attack}

From Eq.(1) we know that the rate of the secure final key is not only
determined by the tagged counts but also determined by the QBER. That is, Eve
may use some other eavesdropping schemes on the multiphoton pulses besides the
PNS attack. Of course, these attacks will cause some QBER which could be
detected in the verification test. A general attack scheme Eve may use is
coherent multiphoton pulses attack. Let us first review the SI attack to
introduce the CMP attack. When a photon propagates from Alice to Bob, Eve can
let a system of her choice, called a probe, interact with the photon. Eve can
freely choose probe and the initial state. But her interaction must obey the
laws of quantum mechanics. That is, her interaction must be described by a
unitary operator. After the interaction, Eve forwards the photon to Bob. Eve
will perform a measurement on her probe to draw Alice's encoding information
after Alice announces the basis she used. This is Eve's SI attack scheme. In
the case of a multiphoton pulse, Eve will let her probes to interact with
Alice's photons one-to-one. After Alice's announcements, Eve will perform a
coherent measurement on her probes. We call this attack as CMP attack.
Obviously, the simplest CMP attack is SI attack: If Alice sends a photon in
the state $|\uparrow\rangle$, the result may be written as
\begin{equation}
U(|\uparrow\rangle|0\rangle)\rightarrow|X\rangle,
\end{equation}
where $|X\rangle$ is the entangled state of the probe and the photon [25].
Likewise, we can obtain the state $|Y\rangle$, $|U\rangle$ and $|V\rangle$
corresponding $|\downarrow\rangle$, $|\rightarrow\rangle$ and $|\leftarrow
\rangle$, respectively. In SI attack scheme, one can obtain that
$|X\rangle=\sqrt{f}|\uparrow\rangle|\phi_{\uparrow}\rangle+\sqrt{e}%
|\downarrow\rangle|\theta_{\uparrow}\rangle$, $|Y\rangle=\sqrt{f}%
|\downarrow\rangle|\phi_{\downarrow}\rangle+\sqrt{e}|\uparrow\rangle
|\theta_{\downarrow}\rangle$, $|U\rangle=\sqrt{f}|\rightarrow\rangle
|\phi_{\rightarrow}\rangle+\sqrt{e}|\leftarrow\rangle|\theta_{\rightarrow
}\rangle$ and $|V\rangle=\sqrt{f}|\leftarrow\rangle|\phi_{\leftarrow}%
\rangle+\sqrt{e}|\rightarrow\rangle|\theta_{\leftarrow}\rangle$, where $f$ is
the fidelity of the state and $f+e=1$. From the unitarity of the interaction,
we have that $\langle\phi_{\uparrow}|\theta_{\uparrow}\rangle=\langle
\phi_{\downarrow}|\theta_{\downarrow}\rangle=\langle\phi_{\uparrow}%
|\theta_{\downarrow}\rangle=\langle\phi_{\downarrow}|\theta_{\uparrow}%
\rangle=0$. It then follows from $\langle\phi_{\uparrow}|\phi_{\downarrow
}\rangle=\cos\alpha$ that QBER=$[1-\cos\alpha]/2$. The maximal information Eve
can gain is that
\begin{equation}
I_{SI}=1-h(\frac{1+2\sqrt{e-e^{2}}}{2}),
\end{equation}
where $h(x)=-x\log_{2}x-(1-x)\log_{2}(1-x)$ and $e$ is QBER.

In Eve's CMP attack scheme, she attaches her probes with all photons in the
multiphoton pulse one-to-one. She interacts the probe-photon pair unitarily
and then forwards the pulse to Bob. She measures the probes coherently after
Alice's announcements. This can be described as%
\begin{equation}
\lbrack U(|\uparrow\rangle|0\rangle)]^{\otimes n}\rightarrow|X\rangle^{\otimes
n},
\end{equation}
where $[U(|\uparrow\rangle|0\rangle)]^{\otimes n}=\underset{n}{\underbrace
{U(|\uparrow\rangle|0\rangle)...U(|\uparrow\rangle|0\rangle)}}$, and
$|X\rangle^{\otimes n}=\underset{n}{\underbrace{|X\rangle...|X\rangle}}$.
Likewise, one can obtain $|Y\rangle^{\otimes n}$, $|U\rangle^{\otimes n}$ and
$|V\rangle^{\otimes n}$. Suppose Alice announces that $|\uparrow\rangle$,
$|\downarrow\rangle$ basis has been used. It has that
\begin{align}
|X\rangle^{\otimes n}  &  =(\sqrt{f}|\uparrow\rangle|\phi_{\uparrow}%
\rangle+\sqrt{e}|\downarrow\rangle|\theta_{\uparrow}\rangle)^{\otimes n},\\
|Y\rangle^{\otimes n}  &  =(\sqrt{f}|\downarrow\rangle|\phi_{\downarrow
}\rangle+\sqrt{e}|\uparrow\rangle|\theta_{\uparrow}\rangle)^{\otimes n}.
\end{align}
Then the two density operators that Eve must distinguish are
\begin{align}
\rho_{\uparrow}  &  =\sum_{i=0}^{n}\frac{n!f^{n-i}e^{i}}{(n-i)!i!}%
|\phi_{\uparrow}\rangle^{\otimes n-i}|\theta_{\uparrow}\rangle^{\otimes
i}(\langle\phi_{\uparrow}|)^{\otimes n-i}(\langle\theta_{\uparrow}|)^{\otimes
i}\\
\rho_{\downarrow}  &  =\sum_{i=0}^{n}\frac{n!f^{n-i}e^{i}}{(n-i)!i!}%
|\phi_{\downarrow}\rangle^{\otimes n-i}|\theta_{\downarrow}\rangle^{\otimes
i}(\langle\phi_{\downarrow}|)^{\otimes n-i}(\langle\theta_{\downarrow
}|)^{\otimes i}%
\end{align}
The optimal information Eve can gain from these two states can be obtained as
follow: Eve first performs the measurements on her probes. If her measurement
results are that $|\phi_{\uparrow}\rangle^{\otimes n-i}|\theta_{\uparrow
}\rangle^{\otimes i}$ (or $|\phi_{\downarrow}\rangle^{\otimes n-i}%
|\theta_{\downarrow}\rangle^{\otimes i}$), where $1\leq i\leq n-1$, then Eve
know that her density operator is $\rho_{\uparrow}$ (or $\rho_{\downarrow}$)
since $\langle\phi_{\uparrow}|\theta_{\uparrow}\rangle=\langle\phi
_{\downarrow}|\theta_{\downarrow}\rangle=\langle\phi_{\uparrow}|\theta
_{\downarrow}\rangle=\langle\phi_{\downarrow}|\theta_{\uparrow}\rangle=0$.
Only if the measurement results are $|\phi_{\uparrow}\rangle^{\otimes n}$,
$|\theta_{\uparrow}\rangle^{\otimes n}$, $|\phi_{\downarrow}\rangle^{\otimes
n}$, and $|\theta_{\downarrow}\rangle^{\otimes n}$, can Eve not distinguish
her density operators. Suppose that Eve's measurement result is $|\phi
_{\uparrow}\rangle^{\otimes n}$. From $\langle\phi_{\uparrow}|\phi
_{\downarrow}\rangle=\cos\alpha$, we can obtain that
\begin{equation}
(\langle\phi_{\downarrow}|\phi_{\uparrow}\rangle)^{\otimes n}=\cos^{n}%
\alpha\text{.}%
\end{equation}
The maximal probability that Eve can distinguish $\rho_{\uparrow}$ from
$\rho_{\downarrow}$ correctly is that $\frac{1+\sqrt{1-\cos^{2n}\alpha}}{2}$.
Thus, the maximal information Eve can gain is that
\begin{align}
I_{CMP}(n)  &  =(1-f^{n}-e^{n})+f^{n}(1-h(\frac{1+\sqrt{1-\cos^{2n}\alpha}}%
{2}))+e^{n}(1-h(\frac{1+\sqrt{1-\cos^{2n}\alpha}}{2}))\nonumber\\
&  =1-(f^{n}+e^{n})h(\frac{1+\sqrt{1-(1-2e)^{2n}}}{2}).
\end{align}
That is, when Eve uses the CMP attack scheme, optimal information she can gain
is $I_{CMP}(n)$. Suppose Eve interacts with $n$ photons. If these $n$ photons
are from $n$ independent qubits (Qubits are uncorrelated since weak coherent
sources are used.), then information Eve can gain is $nI_{SI}$. If these $n$
photons are from a multiphoton pulse, then information Eve can gain is
$I_{CMP}(n)$. When the QBER is small and the photon number $n$ is not so big,
we can gain that $I_{CMP}(n)\geq nI_{SI}$, see Fig.1. In fact, most of the
multiphoton pulses are two-photon pulses since weak coherent sources are used
experimentally. Numerical solution shows that $I_{CMP}(2)>2I_{SI}$ if
$e\leq0.11$, at which error correction can be implemented. That is, CMP attack
is more efficient than SI attack when weak coherent sources are used [26].

\section{from raw key to sifted key}

From discussion above, we know that Eve can get more benefits from a
multiphoton pulse than that from the single-photon pulse. Since Bob's
detection apparatus can resolve the photon number of an arriving pulse, Alice
and Bob can discard all of the multiphoton pulses out of the raw key to
generate the sifted key. Therefore, only the pulses detected in Bob's
detectors as the single photon pulses will be used to generate the sifted key.
In this case, the fraction of counts caused by multiphoton pulses in the
sifted key is that
\begin{align}
\Delta &  =\frac{\sum_{n=2}^{\infty}\mu^{n}e^{-\mu}\eta(1-\eta)^{n-1}%
n/n!}{\sum_{n=1}^{\infty}\mu^{n}e^{-\mu}\eta(1-\eta)^{n-1}n/n!}\nonumber\\
&  =1-e^{-\mu(1-\eta)},
\end{align}
where
\begin{equation}
\lim_{\eta\rightarrow0}\Delta=1-e^{-\mu}.
\end{equation}
That is, the upper bound on the fraction of count caused by multiphoton pulses
is $\Delta_{0}=1-e^{-\mu}$ with high losses. This upper bound is approximate
to $\mu$ when faint coherent sources are used. In order to gain the secure
final key, a fraction $H_{2}(e)$ of the sifted key bits are sacrificed
asymptotically to perform error correction and a fraction $H_{2}(e+\Delta
_{0})$ of the sifted key bits are sacrificed to perform privacy amplification
[27]. After the correcting errors in the sifted key, Alice and Bob can execute
privacy amplification in two different strings, the sifted key bits arising
from the untagged qubits and the sifted key bits arising from the tagged
qubits. The worst case assumption is that the bit error rate is zero for
tagged qubits [7]. Therefore, secure final key can be extracted from sifted
key at the asymptotic rate
\begin{equation}
R\geq(1-\Delta_{0})-H_{2}(e)-(1-\Delta_{0})H_{2}(\frac{e}{1-\Delta_{0}}).
\end{equation}
In the prior art GLLP [7], $\Delta_{0}=p_{multi}/\mu$, where $p_{multi}$ is
the probability of Alice's emitting a multiphoton signal. This is the worst
situation where all the multiphoton pulses mitted by Alice will be received by
Bob. In our scheme, only "single photon" pulses detected in Bob's detectors
are used to generate the sifted key. If this "single photon" pulse is a
multiphoton pulse emitted from Alice, then we assume that it belongs to the
tagged qubits. The other "single photon" pulses detected in Bob's detectors
are real single photon pulses emitted from Alice. Thus, Eve's CMP attack can
be ignored in our scheme.

\section{PDS QKD with imperfect photon-number-resolving detectors}

Resolving power of realistic photon-number-resolving detectors is finite.
Suppose photon number resolving power of the detectors is $n_{0}$. Let us
assume that Eve can attack the photon pulses using PNS attack freely when the
number of a pulses is bigger than $n_{0}$. In this case, additional
information Eve can gain is that
\begin{equation}
\Delta^{\prime}=\frac{\sum_{n=n_{0}+1}^{\infty}\mu^{n}e^{-\mu}/n!}{\sum
_{n=1}^{\infty}\mu^{n}e^{-\mu}\eta(1-\eta)^{n-1}/(n-1)!}.
\end{equation}
Typically, $n_{0}=4$, $\eta=10^{-3}$, $\mu=0.1$. Then we can estimate that
$\Delta^{\prime}\lesssim10^{-3}$, which is a very small quantity. The
particular resolving power of detectors used in Ref.[24] can go up to 10
photons or so ($\sim$8 eV), so that the quantity $\Delta^{\prime}\ll10^{-10}$,
which is negligible. In fact, Eve can not get benefit from the pulses
$n>n_{0}$ since all of the multiphoton pulses detected in Bob's detectors are
discarded, i.e.,$\ \Delta^{\prime}=0$.

Another question is dark counts from blackbody photons propagating through the
optical fiber. Fortunately, these photons can be filtered well.
Experimentally, a really good filter (40 dB out-of-band rejection, 10nm wide
passband), would result in 0.05 Hz of background counts [28]. Suppose the
pulse rate emitted from Alice is $r_{pul}$ and the dark count rate is
$r_{dark}$ Hz. We can obtain the $normalized$ dark count rate $d$ (dark counts
per pulse) in Bob's detectors is that $d\approxeq\frac{r_{dark}}{r_{pul}%
\mu\eta}$. Distribution of dark counts in Bob's detectors is that
\begin{equation}
p_{dark}[n]=d^{n}.
\end{equation}
Therefore, in experiment, Bob can obtain photon number distribution of the
laser pulse by subtracting the dark counts from the real counts.\ QBER
$e_{dark}$ caused by dark counts should be considered, especially in the long
distance QKD,%
\begin{equation}
e=e_{0}+e_{dark},
\end{equation}
where $e_{dark}=d/2$, and $e_{0}$ is caused by the imperfections of the
optical setup [1].

\section{optimal intensity of laser source to generate secure final key}

In BB84, the rate of generating raw key is approximate to $\frac{1}{4}\mu\eta
$. Thus, the rate of generating secure final key is approximate to $\frac
{1}{4}\mu(1-\Delta_{0})\eta R$. That is, the rate of generating the secure
final key is approximate to $R_{f}$, where
\begin{equation}
R_{f}=\frac{1}{4}\mu(1-\Delta_{0})\eta\lbrack(1-\Delta_{0}-H_{2}%
(e)-(1-\Delta_{0})H_{2}(\frac{e}{1-\Delta_{0}})],
\end{equation}
where $\Delta_{0}=1-e^{-\mu}$. In practice, $e$ and $\eta$ are constants when
the transmission distance is constant. Therefore, the only variable in $R_{f}$
is $\mu$. $R_{f}$ reaches its maximum at the point $\frac{\partial R_{f}%
}{\partial\mu}=0$. In this way, we can obtain optimal parameter $\mu$, see
Fig. 2.

\section{discussion and conclusion}

In the prior decoy state QKD [14,15,17], it requires that $\mu^{\prime}>\mu$.
In [14,15], the upper bound on the fraction of counts caused by the
multiphoton is $\Delta\leq\frac{\mu e^{-\mu}}{\mu^{\prime}e^{-\mu^{\prime}}}$.
Only if $\mu=\mu^{\prime}$ can the upper bound be reduced to $\mu$ [15]. In
our scheme, $\mu$ is independent of $\mu^{\prime}$ so that both signal pulse
and decoy pulses can be used to generate the raw key. Another difference is
that all the pulses detected in Bob's detectors are discarded in our scheme,
so that Eve's CMP attack does not exist in our scheme. However, CMP attack
should be considered in [14,15,17] to ensure the security of the final key.

In our scheme, from $\Delta=1-e^{-\mu(1-\eta)}$, we can conclude that the
upper bound $\Delta_{0}=1-e^{-\mu}$ can not be reduced any longer as long as
weak coherent sources and high lossy channel are used, so that the quantity
$\Delta_{0}=1-e^{-\mu}$ is also the lower bound on the fraction of counts
caused by the multiphoton pulses. Thus, the fraction $\Delta_{0}=1-e^{-\mu}$
seems \textquotedblleft$inherent$\textquotedblright\ in the long distance QKD
with weak coherent sources and high lossy channel.

In summary, we have discussed the security of practical BB84 QKD protocol with
weak coherent sources, noises and high losses. We have presented a PDS QKD
scheme based on recent experimental advancements. The upper bound on fraction
of counts caused by multiphoton pulses is independent of the intensity of
decoy source so that both the signal pulses and decoy pulses can be
implemented to generate the raw key after verified the security of the QKD. We
have shown that CMP attack is more efficient than SI attack. Finally, optimal
$\mu$ is presented to improve the rate of generating the secure final key.

\section{acknowledgment}

We are grateful to D. Rosenberg for his help. This work is supported by
National Natural Science Foundation of China under Grant No. 10447140 and 10504039.

\section{references}

[1] N. Gisin, G. Ribordy, W. Tittel, and H. Zbinden, Rev. Mod. Phys.
\textbf{74}, 145-195 (2002).

[2] C. H. Bennett, and G. Brassard, in\textit{ Proceedings of the IEEE
International Conference on Computers, Systems and Signal Processing,}
Bangalore, India, (IEEE, New York, 1984), pp.175-179.\smallskip

[3] D. Mayers, J. of ACM \textbf{48}, 351 (2001).

[4] E. Biham, M. Boyer, P. O. Boykin, T. Mor, and V. Roychowdhury, in\textit{
Proc. of the thirty-second annual ACM symposium on Theory of computing}
(Portland,Oregon, United States,2000), pp. 715--724.

[5] H.-K. Lo and H. F. Chau, Science \textbf{283}, 2050(1999).

[6] P. W. Shor and J. Preskill, Phys. Rev. Lett. \textbf{85}, 441 (2000).

[7] D. Gottesman, H.-K. Lo, N.L\"{u}kenhaus, and J. Preskill, Quant. Inf.
Comp. \textbf{5}, 325(2004).

[8] C. H. Bennett, F. Bessette, G. Brassward, L. Salvail, and J. Smolin, J.
Cryptology \textbf{5}, 3-28 (1992); For a review, please see Ref.[1] and
references theirn.

[9] B. Huttner, N. Imoto, N. Gisin, and T. Mor, Phys. Rev. A \textbf{51}, 1863 (1995).

[10] H. P. Yuen, Quantum Semiclassical Opt. \textbf{8}, 939 (1996).

[11] N. L\"{u}kenhaus and M. Jahma, New J. Phys. \textbf{4}, 44 (2002).

[12] N. L\"{u}kenhaus, Phys. Rev. A 61, 052304 (2000).

[13] G. Brassard, N. L\"{u}kenhaus, T. Mor, and B. C. Sanders, Phys. Rev.
Lett. \textbf{85}, 1330(2000).

\bigskip\lbrack14] W.-Y. Hwang, Phys. Rev. Lett. \textbf{91}, 057901 (2003).

[15] X.-B. Wang, Phys. Rev. Lett. \textbf{94}, 230503 (2005).

[16] X.-B. Wang, Phys. Rev. A \textbf{72}, 012322 (2005).

[17] H.-K. Lo, X. Ma, and K. Chen, Phys. Rev. Lett. \textbf{94}, 230504 (2005).

[18] H.-K. Lo, in\textit{ Proc. of IEEE International Symposium on Information
Theory (ISIT)} 2004 (2004), p.137.

[19] X. Ma, B. Qi, Y. Zhao, and H.-K. Lo, Phys. Rev. A \textbf{72}, 012326 (2005).

[20] J. W. Harrington, J. M. Ettinger, R. J. Hughes, and J. E. Nordholt,
(2005) arXiv:quant-ph/0503002.

[21] V. Scarani, A. Acin, G. Ribordy, and N. Gisin, Phys. Rev. Lett.
\textbf{92}, 057901 (2004); C. Branciard, N. Gisin, B. Kraus, and V. Scarani,
Phys. Rev. A \textbf{72}, 032301 (2005); Chi-Hang F. Fung, K. Tamaki, and
H.-K. Lo, (2005) arXiv:quant-ph/0510025.

\bigskip\lbrack22] D. Achilles, C. Silberhorn, C. Sliwa, K. Banaszek, I. A.
Walmsley, M. J. Fitch, B. C. Jacobs, T. B. Pittman, and J. D. Franson, J. Mod.
Opt. \textbf{51}, 1499 (2004).

[23] E. Waks, K. Inoue, W. D. Oliver, E. Diamanti, and Y. Yama-moto, IEEE J.
Sel. Top. Quantum Electron. \textbf{9}, 1502 (2003).

[24] D. Rosenberg, A. E. Lita, A. J. Miller, and S. W. Nam, Phys. Rev. A
\textbf{71},061803(R) (2005).

[25] C. A. Fuchs, N. Gisin, R. B. Griffiths, C.-S. Niu, and A. Pere, Phys.
Rev. A \textbf{56}, 1163-1172 (1997).

[26] Some correlative works can be found in: M. Curty and L\"{u}tkenhaus,
Phys. Rev. A \textbf{69}, 042321 (2004); A. Niederberger, V. Scarani, N.
Gisin, Phys. Rev. A \textbf{71}, 042316 (2005).

[27] P. Shor and J. Preskill, Phys. Rev. Lett. \textbf{85}, 441-444 (2000).

[28] Maybe, we should assume that Eve can control the dark counts since Eve
may change the wavelength of Alice's photon which is more sensitive for Bob's
detectors. However, Bob can adds a filter in his laboratory to defeat Eve's
such attacks. These days, the bandwidth of optical devices is as narrows as
0.1 to 0.01nm which is comparable to the laser linewidth. An optical grating
to filter out unwanted frequencies may be used in combination with such the
narrow bandwidth devices; Experimental data were obtained from D. Rosenberg by
private communication.

[29] C. Gobby, Z. L. Yuan, and A. J. Shields, Appl. Phys. Lett. \textbf{19},
3762 (2004).

\section{caption}

Caption 1. (Color online.) Information vs photon number. Information Eve can
gain from $n$ photons by using SI attack (a) is $nI_{SI}$ since these $n$
photons come from $n$ uncorrelated photon pulses. If these $n$ photons are
from a multiphoton pulse, then information Eve can gain is $I_{CMP}(n)$ (b).
Numerical solution shows that $I_{CMP}(2)>2I_{SI}$ when $e\leq11\%$. And
$I_{CMP}(3)>3I_{SI}$ when $e\leq6.8\%$. CMP attack is more efficient than SI
attack since weak coherent sources are used experimentally.

Caption 2. (Color online.) Rate of generating final key vs transmission
distance. In order to be comparable, we use the parameters in [17,29] instead
of [24]. When $\mu=0.1$, transmission distance is close to 140.2 km which is
comparable with LMC in [17]. Numerical solution shows that optimal intensity
of laser source is $\mu\thickapprox0.7$ (transmission distance over 164.1 km).
That is, optimal intensity of laser source provides 23.9 km increase in the
transmission distance. Transmission distance is stable to small perturbations
to the optimal $\mu$ (up to 20\% change of $\mu$, less than 0.3\% change of
transmission distance). Here, we have verified that error correction are
allowable for the maximal transmission distance.

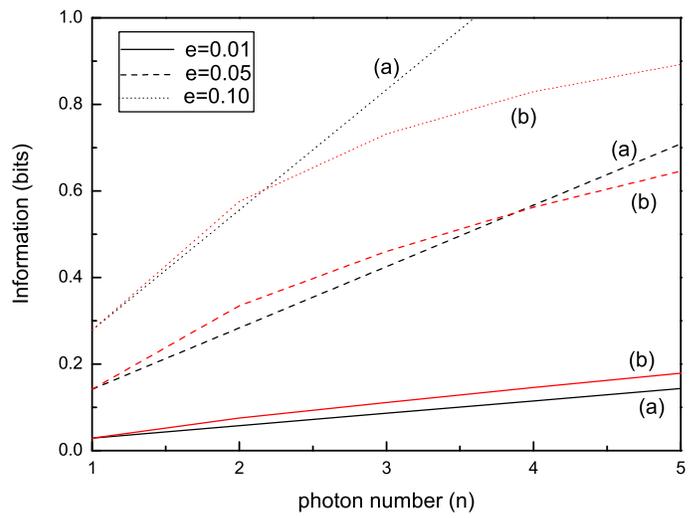
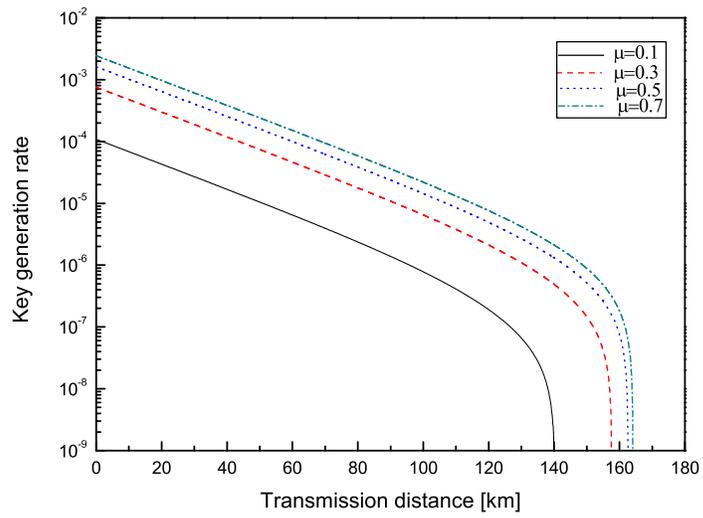
\end{document}